\documentclass[%
 reprint,
 amsmath,amssymb,
 aps,
prb,
superscriptaddress,
longbibliography
]{revtex4-2}

\usepackage{graphicx}
\graphicspath{{./Images/}}
\usepackage{dcolumn}
\usepackage{bm}

\begin{document}

\title{Energy distribution controlled ballistic Josephson junction}

\author{P. Pandey}
\email{preeti.pandey@aalto.fi}
\affiliation{
Low Temperature Laboratory, QTF Centre of Excellence, Department of Applied Physics, Aalto University School
of Science, P.O. Box 15100, 00076 Aalto, Finland.
}

\author{D. Beckmann}
 \email{detlef.beckmann@kit.edu}
\affiliation{ 
Institute for Quantum Materials and Technologies, Karlsruhe Institute of Technology, \\Karlsruhe D-76021, Germany
}

\author{R. Danneau}
 \email{romain.danneau@kit.edu}
\affiliation{ 
Institute for Quantum Materials and Technologies, Karlsruhe Institute of Technology, \\Karlsruhe D-76021, Germany
}

\begin{abstract}
We report an experimental study on the tuning of supercurrent in a ballistic graphene-based Josephson junction by applying a control voltage to a transverse normal channel. In this four-terminal geometry, the control voltage changes the occupation of Andreev states in the Josephson junction, thereby tuning the magnitude of the supercurrent. As a function of gate voltage, we find two different regimes characterized by a double-step distribution and a hot-electron distribution, respectively. Our work opens new opportunities to design highly controllable Josephson junctions for tunable superconducting quantum circuits.  
\end{abstract}

\maketitle

\section{Introduction}
The design of controlled Josephson junctions is a crucial  aspect of engineering superconducting circuits and their potential application in quantum technologies and information \cite{krantz2019}. This is particularly important in Josephson junctions based on superconducting weak links \cite{likharev1979}. Many different systems have been used to mediate superconductivity induced by proximity of superconducting leads such as metallic \cite{vandover1981,morpurgo98,baselmans1999,baselmans2001,dubos2001} or semiconducting \cite{clark1980,volkov1995,bastian1998,schapersbook,schapers2003,schapers2003a,wan2015,shabani2016} nanostructures, quantum point contacts \cite{amado2013,kraft2018}, quantum dot \cite{defranceschi2010}, topological insulators \cite{sacepe2011,williams2012,veldhorst2012,hart2014,bocquillon2017,dartiailh2021,schmitt2022} or graphene \cite{lee2018}. When superconducting leads connecting the weak link are close enough to each other, a dissipationless current can flow via Andreev bound states \cite{kulik1970}. Dedicated low noise electronics and low temperature filtering is required to measure finite \cite{steinbach2001,wu2009,maurand2012,thalmann2017} and stable \cite{Haque2021} maximum supercurrent or critical current $I_{\mathrm{c}}$. A full control strategy strongly depends on the tunability of the superconducting weak link itself. Indeed, $I_{\mathrm{c}}$ depends on many parameters such as temperature, superconducting gap, contact transparency or/and disorder. 

Moreover, when the weak link carrier density is gate tunable, the critical current amplitude varies inversely to the normal state resistance $R_{\mathrm{N}}$, i.e. $I_{\mathrm{c}}$ increases when $R_{\mathrm{N}}$ decreases. Therefore, the electrostatic control remains the most straight forward way to tune the amplitude of the critical current \cite{clark1980,wan2015,shabani2016,sacepe2011,williams2012,veldhorst2012,hart2014,bocquillon2017,dartiailh2021,schmitt2022,lee2018}, while local gating allows its spatial confinement \cite{amado2013,kraft2018,defranceschi2010}. An alternative way to control the amplitude of $I_{\mathrm{c}}$ was demonstrated by tuning the electron distribution in the weak link itself with a transverse controlled voltage \cite{morpurgo98,baselmans1999,baselmans2001,schapers2003,schapers2003a}. In these experiments, the control voltage allowed to tune $I_{\mathrm{c}}$ up to reversing its sign and therefore forming a so-called $\pi $-junction \cite{baselmans1999,baselmans2001}. In this work, we study the effect of such control voltage in a ballistic superconducting weak link, here a single layer graphene, in which the supercurrent is no longer carried by a continuum of Andreev bound states (ABS) but instead a series of discrete states \cite{kulik1970}. We show that this transverse voltage tunes the occupation of Andreev states in the Josephson junction, adding an additional knob to control the magnitude of the supercurrent. We have used a model accounting for the coupling of the control channel to the Josephson junction  \cite{bagwell1992,chang1997,ilhan1998,melin2022} with the transparency of the junction as a single parameter. We obtain two distinct regimes, a double-step distribution and a hot-electron distribution, which fits the amplitude of $I_{\mathrm{c}}$ when the gate voltage tunes the Fermi level in the valence and the conduction band, respectively, corresponding to a high and low normal state resistance.

\begin{figure}
\includegraphics[width=0.5\textwidth]{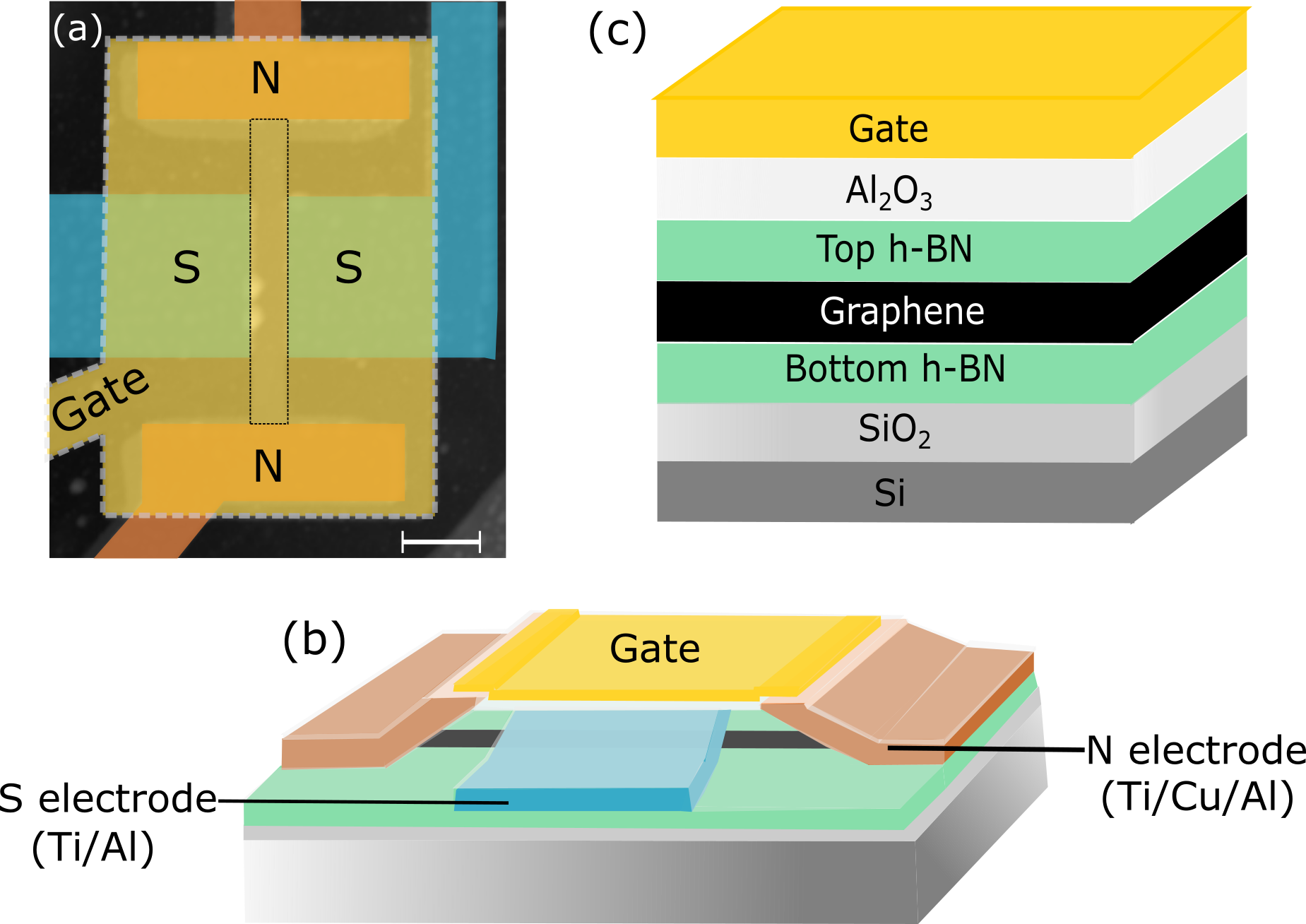}
\caption{(a) False color atomic force micrograph of the studied device showing the normal metal (N), superconducting (S), and top gate (Gate) electrodes. The region enclosed by the dashed black lines shows the h-BN encapsulated graphene channel. The scalebar is 1\,$\mu$m. (b) Schematics of the device depicting the edge contact geometry. (c) Cross-section schematic of the central part of the device.}
\label{fig:Figure1}
\end{figure}

\section{Experimental results}
\subsection{Device fabrication and measurement setup}

We have fabricated our device by encapsulating a single layer graphene between two h-BN crystallites, 28 and 18\, nm thick for the top and bottom h-BN, respectively. Graphene and h-BN crystallites were mechanically exfoliated from natural graphite (NGS Naturgraphit GmbH) and commercially available h-BN powder (Momentive, gradePT110), respectively,  on a Si substrate with 300\,nm thick SiO$_{2}$. The h-BN/graphene/h-BN heterostructure was prepared by following a dry transfer technique similar to Wang \textit{et.\,al.}\,\cite{Wang2013}. The fabricated heterostructure was subsequently transferred to a different Si substrate with 1000\,nm thick SiO$_{2}$. Edge contacts to the h-BN encapsulated graphene were established in a similar manner as in Kraft \textit{et.\,al.}\,\cite{kraft2018}. However, this process was adapted for the deposition of two different types of metal electrodes as in Pandey \textit{et.\,al.}\,\cite{Pandey2019}. For the superconducting electrodes, we have used Ti(5\,nm)/Al(60\,nm), and for the normal metal electrodes, Ti(5\,nm)/Cu(80\,nm)/Al(5\,nm) was used. To control the charge carrier density in the graphene channel, a top gate electrode with Ti(5\,nm)/Cu(100\,nm)/Al(5\,nm) was fabricated. Here, a 25\,nm thick Al$_{2}$O$_{3}$ was deposited with atomic layer deposition method to act as the top gate dielectric in addition to the top h-BN. In all the electrodes, Ti serves as an adhesive layer, while the thin Al layer in the Cu contact and gate electrodes is used as a protective layer to prevent oxidation of these electrodes.

As can be see in Fig.\,\ref{fig:Figure1}(a), the graphene channel has $L=0.46\,\mu$m and $W=2\,\mu$m across the SGS junction and $L=4\,\mu$m and $W=0.46\,\mu$m across the NGN junction. Note that $L$ denotes the length of the channel between the electrodes, while $W$ denotes the width of the channel along the electrodes. Fig.\,\ref{fig:Figure1}(b) shows the schematic of the device with the edge contact geometry and Fig.\,\ref{fig:Figure1}(c) shows the cross-section schematic of the device. All of the transport measurements were conducted with standard low frequency lock-in technique. In order to avoid any spurious effects, the measurement lines were fed through a series of filters. In all of the data shown, $I_\mathrm{bias}$($V_\mathrm{bias}$) denotes the current(voltage) across the SGS junction, whereas $V_\mathrm{ctrl}$ denotes the voltage bias across the NGN junction.

\begin{figure}
\includegraphics[width=0.4\textwidth]{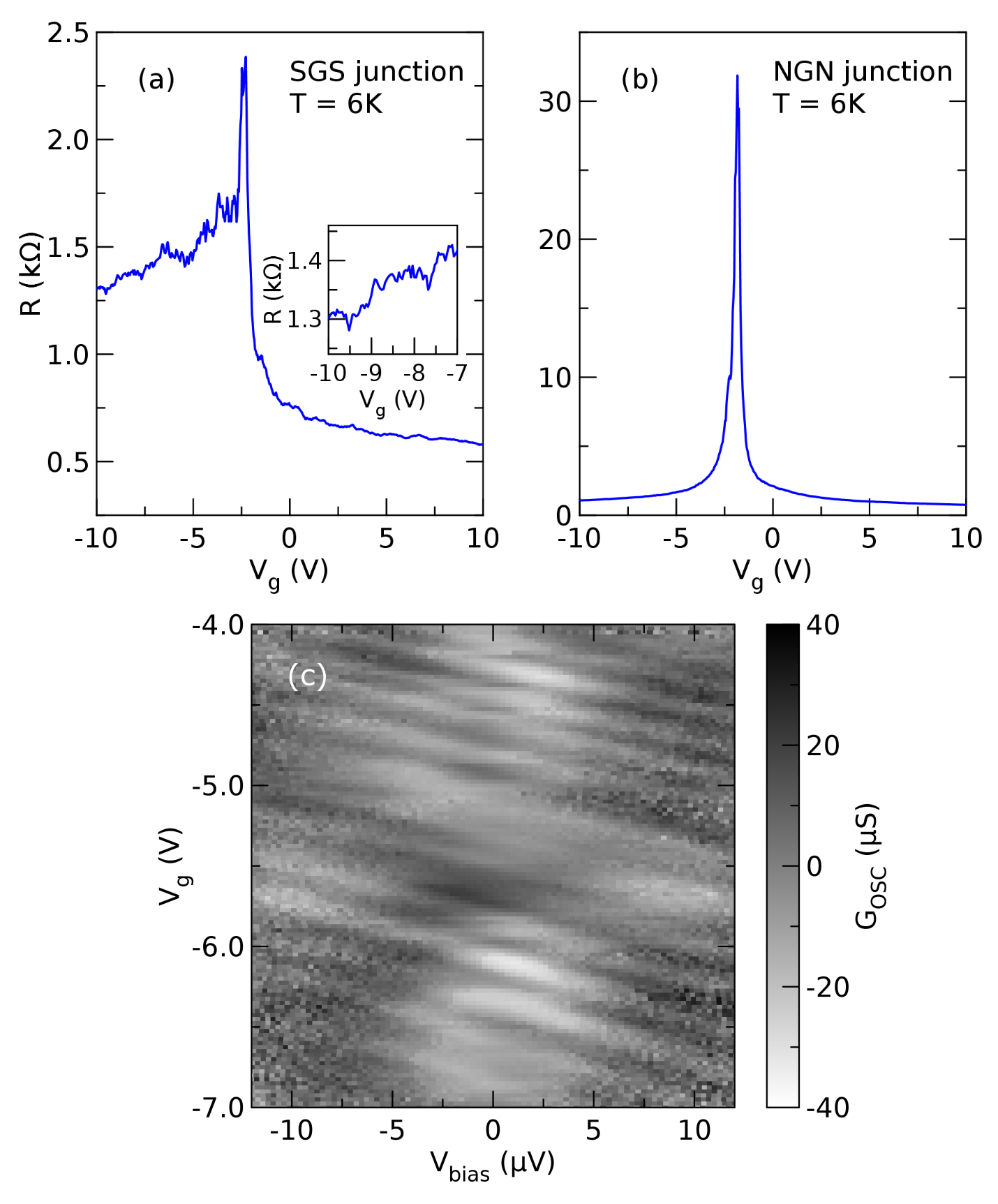}
\caption{Resistance $R$ of the device as a function of gate voltage $V_\mathrm{g}$ as measured across: (a) Superconductor/graphene/superconductor (SGS) junction, the inset shows Fabry-P\'{e}rot resonances in the p-doped regime, (b) Normal metal/graphene/normal metal (NGN) junction. (c) Fabry-P\'{e}rot oscillations in the SGS junction as a function of gate voltage $V_\mathrm{g}$ and bias voltage $V_\mathrm{bias}$ across the SGS junction.}
\label{fig:Figure2}
\end{figure}

\subsection{Normal state characterization and ballistic transport}

Fig.\,\ref{fig:Figure2}(a) and (b) show the gate voltage $V_\mathrm{g}$ dependent resistance of the device across the SGS and NGN junctions, respectively, when the device is in the normal state with $T=6\,$K. It can be readily observed that the Dirac point for both of these junctions is shifted to negative gate voltages which indicates n-type doping of the graphene sheet due to charge transfer from the contacts. It also results in the formation of potential barriers close to the metal/graphene contact interfaces when the charge transport in the graphene channel is driven into the p-doped regime \cite{Giovannetti2008,Khomyakov2009,Khomyakov2010}. An important point to note is that the two junctions have different Dirac points where $V_\mathrm{CNP}=-2.3\,$V for the SGS junction and $V_\mathrm{CNP}=-1.8\,$V for the NGN junction. In addition, comparing the charge transport between the n-doped and p-doped regions, higher asymmetry is observed in the SGS junction while the Dirac curve looks symmetric across the NGN junction. This can be explained by considering the dimensions of the two junctions, where the SGS junction falls in the category of wide and short junction. NGN junction, on the other hand, falls in the narrow and long junction regime. Therefore, the doping from the contacts is more prominent in the SGS junction leading to a higher shift in the $V_\mathrm{CNP}$ and charge transport asymmetry in the Dirac curve as compared to the NGN junction. It also results in the formation of a coherent electronic Fabry-P\'{e}rot (FP) cavity, a signature of ballistic transport, in the SGS junction when the Fermi level in graphene is tuned in the valence band \cite{katsnelsonbook,Miao2007,Shytov2008,Young2009,Cho2011,Grushina2013,Rickhaus2013,Varlet2014,Du2018,kraft2021}. 

This effect can be readily seen in the inset of Fig.\,\ref{fig:Figure2}(a) as the conductance oscillates with the gate voltage. These conductance oscillations can be tuned by $V_\mathrm{g}$ as well as by an applied bias $V_\mathrm{bias}$ across the SGS junction. Fig.\,\ref{fig:Figure2} (c) shows these conductance oscillations mapped with respect to $V_\mathrm{g}$ and $V_\mathrm{bias}$. As the conductance of the device varies strongly with $V_\mathrm{g}$ and $V_\mathrm{bias}$, a non-oscillating background conductance was subtracted to enhance the visibility of the conductance oscillations. As can be seen, instead of the usual checkerboard pattern for symmetric contacts, we observe an asymmetric interference pattern which suggest an asymmetric coupling of the two S contacts to the graphene channel \cite{Wu2007,Pandey2019}. Despite this asymmetric coupling clear interferences can be observed in Fig.\,\ref{fig:Figure2} (c). For normal incidence of charge carriers, the cavity length $L_\mathrm{C}$ that gives rise to this pattern can be calculated by using the relation $k_\mathrm{f}L_\mathrm{C}=N\pi$ where $N$ is an integer and $k_\mathrm{f}=\sqrt{n\pi}$ is the Fermi wavevector with $n=\alpha_\mathrm{g}V_\mathrm{g}$ denoting the charge carrier density. Here $\alpha_\mathrm{g}=3.496\times10^{11}$\,V$^{-1}$cm$^{-2}$ is the gate coupling efficiency which was extracted from Shubnikov-de Haas oscillations at high field (not shown). Inserting $\alpha_\mathrm{g}$ in the expression to calculate $L_\mathrm{C}$ results in $L_\mathrm{C}\sim\,450\pm\,5$\,nm. Since $L_\mathrm{C}$ is in agreement with the length of the graphene channel across the SGS junction, it can be concluded that the observed interference pattern is indeed arising from the FP oscillations. This serves as the proof that the charge transport across the SGS junction is in the ballistic regime.

\begin{figure}
\includegraphics[width=0.5\textwidth]{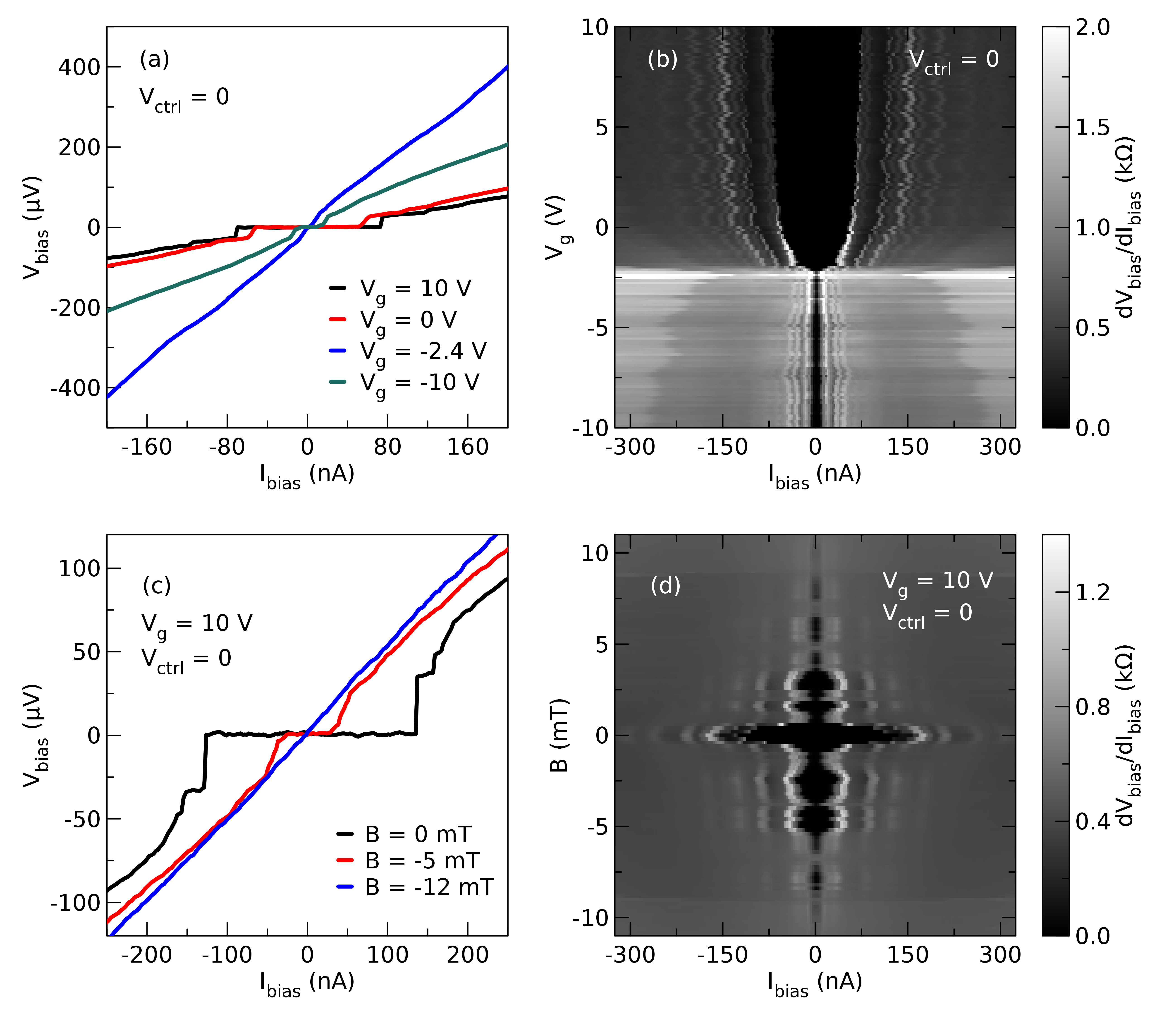}
\caption{Characterization of the supercurrent at $T=50$\,mK and $V_\mathrm{ctrl}=0$. (a) $I_\mathrm{bias}-V_\mathrm{bias}$ characteristics at different gate voltages $V_\mathrm{g}$. (b) Gate voltage dependence of the differential resistance $dV_\mathrm{bias}/I_\mathrm{bias}$. (c) $I_\mathrm{bias}-V_\mathrm{bias}$ characteristics in the presence of perpendicular magnetic field $B$ at $V_\mathrm{g} = 10$\,V. (d) Measured Fraunhofer pattern ($B$ vs $I_\mathrm{bias}$) under the same conditions as in (c). Please note that a magnetic field offset correction was made for (c) and (d).}
\label{fig:Figure3}
\end{figure}

\subsection{Gate dependent supercurrent and Fraunhofer pattern}

Here, we characterize the SGS Josephson junction in the absence of a control voltage across the NGN channel at $T=50~\mathrm{mK}$, far below the critical temperature of Al ($\sim\,1$\,K).
Supercurrent in ballistic graphene based Josephson junctions can be efficiently tuned by an applied gate voltage which tunes the charge carrier density in the weak link \cite{Calado2015,BenShalom2016,Borzenets2016,kraft2018,pandey2021}. Fig.\,\ref{fig:Figure3}(a) shows the $I_\mathrm{bias}-V_\mathrm{bias}$ characteristics of the SGS junction. It is to be noted that due to a slight shift in the zero magnetic field, the device was in a nonzero magnetic field during this set of measurements which resulted in a reduced supercurrent across the SGS junction. The $I_\mathrm{bias}-V_\mathrm{bias}$ curves represent the high p-doped regime ($V_\mathrm{g}=-10$\,V), the vicinity of the Dirac point ($V_\mathrm{g}=-2.4$\,V), and the n-doped regime ($V_\mathrm{g}=0$\,V and $V_\mathrm{g}=10$\,V). Fig.\,\ref{fig:Figure3}(b) shows the complete map of the differential resistance $dV_\mathrm{bias}/dI_\mathrm{bias}$ across the SGS junction as a function of $I_\mathrm{bias}$ and $V_\mathrm{g}$ under the same measurement conditions as in Fig.\,\ref{fig:Figure3}(a). It can be clearly seen that the device has a noticeable supercurrent ($\sim\,80$\,nA) in the entire n-doped regime while it is considerably smaller even in the highly p-doped regime. Close to the Dirac point, the device is very close to the normal state with slight non-linearity in the current-voltage relation. As can be seen in Fig.\,\ref{fig:Figure3}, the gate tuning of the supercurrent in our device agrees with the commonly observed gate-tunable supercurrent in graphene based Josephson junctions \cite{Calado2015,BenShalom2016,Borzenets2016}.

\begin{figure*}
\includegraphics[width=1\textwidth]{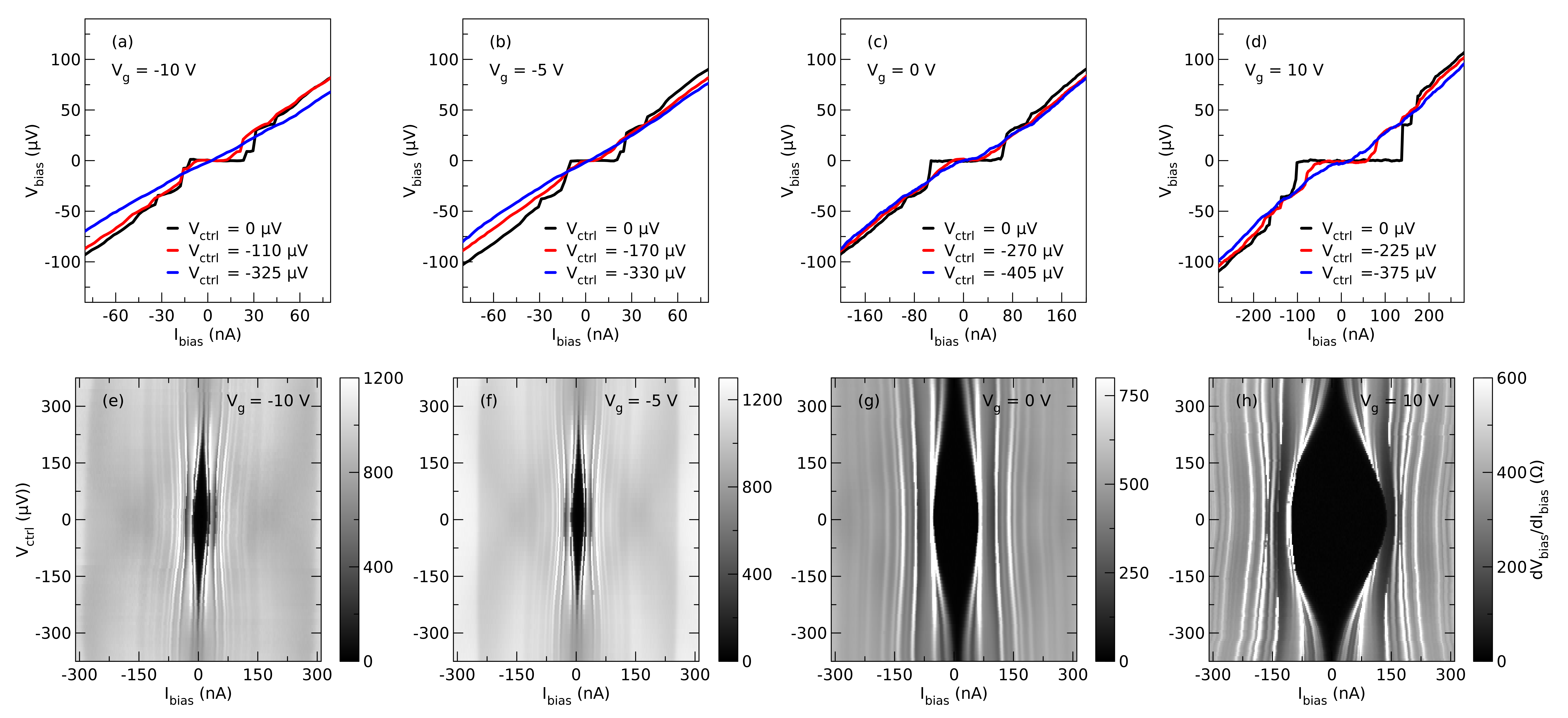}
\caption{$I_\mathrm{bias}-V_\mathrm{bias}$ characteristics for different control voltages $V_\mathrm{ctrl}$ at (a) $V_\mathrm{g} =-10$\,V, (b) $V_\mathrm{g} =-5$\,V, (c) $V_\mathrm{g} =0$\,V, (d) $V_\mathrm{g} =10$\,V. Differential resistance $dV_\mathrm{bias}/dI_\mathrm{bias}$ map as a function of $I_\mathrm{bias}$ and $V_\mathrm{ctrl}$ at (e) $V_\mathrm{g} =-10$\,V, (f) $V_\mathrm{g} =-5$\,V, (g) $V_\mathrm{g} =0$\,V, (h) $V_\mathrm{g} =10$\,V. All data were taken at $T=50$\,mK and $B=0$.}
\label{fig:Figure4}
\end{figure*}

In a wide and short Josephson junction with uniform current density, the critical current is expected to exhibit a Fraunhofer interference pattern according to the relation  \cite{BaronePaternoBook}
\begin{equation}
I_\mathrm{c}(B) = I_\mathrm{c}(0)\,\frac{\sin(\pi\Phi/\Phi_{0})}{\pi\Phi/\Phi_{0}}
\label{eq:B_dep_Ic}
\end{equation}
Here $I_\mathrm{c}(B)$ is the critical current at a perpendicular magnetic field $B$, $I_\mathrm{c}(0)$ is the critical current at zero magnetic field, $\Phi = B A$ is the magnetic flux through the Josephson junction with junction area $A$, and $\Phi_{0}$ is the magnetic flux quantum. The Fraunhofer pattern bears a strong dependence on the current density distribution in the Josephson junction. Fig.\,\ref{fig:Figure3}(c) and (d) show the $I_\mathrm{bias}-V_\mathrm{bias}$ characteristics and the differential resistance, respectively, in the presence of a perpendicular magnetic field $B$ (offset-corrected), at $V_\mathrm{g} = 10$\,V. We observe a maximum critical current of $\sim$130\,nA at zero magnetic field and an oscillatory pattern as a function of field is seen in Fig.\,\ref{fig:Figure3}(d). A clear asymmetry in the interference pattern for positive and negative magnetic fields is due to the flux trapping in the magnet itself which could not be corrected in the measurement and analysis. The interference pattern resembles a Fraunhofer pattern, and the field scale of the oscillation is consistent with the expected $\Delta B\approx 2~\mathrm{mT}$ for the geometry of our device.

\subsection{Tuning the supercurrent with a transverse normal channel}

Following the characterization of the Josephson junction, we now turn to the control of the occupation of Andreev states in the SGS channel by applying a control voltage $V_\mathrm{ctrl}$ across the NGN channel. Fig.\,\ref{fig:Figure4}(a)-(d) show the $I_\mathrm{bias}-V_\mathrm{bias}$ curves across the SGS junctions at various values of $V_\mathrm{ctrl}$ at 50\,mK under zero magnetic field. The charge transport in the p-doped regime corresponds to Fig.\,\ref{fig:Figure4}(a) and (b), while Fig.\,\ref{fig:Figure4}(c) and (d) correspond to the n-doped regime. Fig.\,\ref{fig:Figure4}(e)-(h) show the complete map of $dV_\mathrm{bias}/dI_\mathrm{bias}$ across the SGS junction with respect to $I_\mathrm{bias}$ and $V_\mathrm{ctrl}$ under the same measurement conditions as in Fig.\,\ref{fig:Figure4}(a)-(d). The asymmetry with respect to the positive and negative $I_\mathrm{bias}$ in Fig.\,\ref{fig:Figure4}(d) and (h) can be attributed to the self-heating of the device at high critical current. While the supercurrent is continuously tuned by the control voltage in all of the measurements, a clear trend can be observed in all of the maps that is the magnitude of the control voltage required to diminish the superconductivity depends on the magnitude of the critical current. A comparatively smaller control voltage can clearly suppress the supercurrent in the p-doped regime as seen in Fig.\,\ref{fig:Figure4}(a),(b),(e) and (f), while the supercurrent cannot be completely suppressed in the n-doped regime as observed in Fig.\,\ref{fig:Figure4}(c),(d),(g) and (h) with the observable non-linearity even at high control voltages. It can be understood by considering the fact that the device is weakly superconducting in the p-doped regime as compared to the n-doped regime, and therefore, it is more sensitive to the change caused by the control voltage in the p-doped regime.

\begin{figure*}
\includegraphics[width=\textwidth]{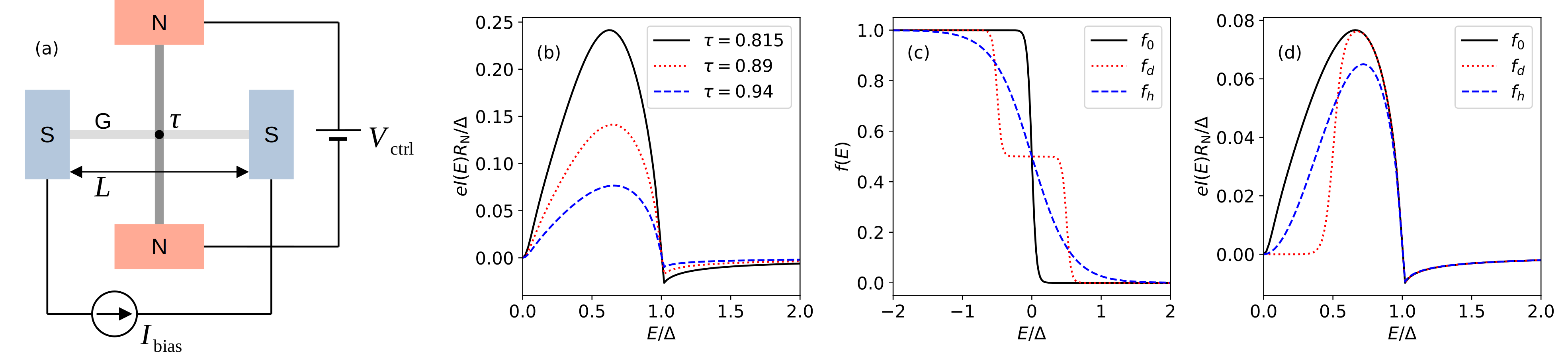}
\caption{(a) Schematic view of the model. (b) Spectral supercurrent at the critical current $I_\mathrm{c}$ for $V_\mathrm{ctrl}=0$ for three different transparencies $\tau$. (c) Distribution functions $f_0$, $f_d$ and $f_h$ for $eV_\mathrm{ctrl}=\Delta$. (d) Spectral supercurrent for $\tau=0.94$ and the same distribution functions as in (c).}
\label{fig:Figure5}
\end{figure*}

\section{Model}

Supercurrent is carried across a ballistic Josephson junction by discrete Andreev bound states (ABS) below the gap and additional continuum contributions from "leaky" Andreev resonances above the gap \cite{kulik1970,bagwell1992}. The magnitude and direction of the supercurrent is controlled by the phase difference across the junction and the occupation of the states. In our case, the Josephson junction is strongly coupled to the control channel, since they are both part of the same graphene sheet. This coupling leads to dephasing of the Andreev levels with a concomitant reduction of the critical current. We use the model of a strongly coupled control channel \cite{chang1997} generalized to the four-terminal geometry of our experiment \cite{ilhan1998}. The model is illustrated in Fig.~\ref{fig:Figure5}(a). Both the SGS Josephson junction and the NGN control channel are modeled as single-channel one dimensional conductors, coupled by a junction of transparency $\tau$. The length $L=460~\mathrm{nm}$ of the SGS junction was taken from the geometry of the device. The coherence length of the Andreev pairs in the graphene sheet is $\xi_G=\hbar v_\mathrm{F}/2\Delta=2.4~\mathrm{\mu m}$, where $v_\mathrm{F}=10^6~\mathrm{m/s}$ is the Fermi velocity of graphene, and $\Delta=135~\mathrm{\mu eV}$ is the experimentally observed gap of the aluminum leads. Therefore, our junction is in the short limit, and we expect a single ABS below the gap. As an illustration, we show the spectral supercurrent $I(E)$ at the critical current for three different values of $\tau$ in Fig.~\ref{fig:Figure5}(b). As can be seen, the single ABS is significantly broadened by the coupling to the control channel. In addition, there is a small negative contribution from the continuum of "leaky" states above the gap.

The supercurrent is controlled by the applied voltage $V_\mathrm{ctrl}$. The distribution functions for ideal reservoirs are given by
\begin{equation}
	f_\pm(E) = f_0\left(E\pm\frac{eV_\mathrm{ctrl}}{2}, T\right),
\end{equation}
where $f_0$ is the Fermi distribution, and $\pm$ corresponds to the upper/lower control terminal. In the absence of inelastic scattering, the distribution function in the middle of the control wire is the superposition
\begin{equation}
	f_\mathrm{d}(E) = \frac{1}{2}\left[f_+\left(E\right) + f_-\left(E\right)\right],
\end{equation}
{\em i.e.}, a double-step distribution function \cite{pothier1997}. The double-step distribution assumes perfect cooling of the dissipated electric power by the reservoirs. In the n-doped regime, the lower resistance of the graphene sheet leads to increased power dissipation and self-heating of the device, as already seen in the asymmetry of the switching and retrapping currents in Fig.~\ref{fig:Figure4} \cite{courtois2008}. As an alternative scenario, we therefore also consider the limiting case of complete thermalization by Coulomb scattering, {\em i.e.}, the hot-electron regime. In this case, the double-step distribution is replaced by a  distribution
\begin{equation}
f_\mathrm{h}(E) = f_0(E, T^*).
\end{equation}
with an increased effective temperature $T^*$. Here, we neglect electron-phonon cooling, which leaves only cooling by electronic heat transport to the reservoirs. In this case, for a voltage-biased wire, the Wiedemann-Franz law gives the temperature \cite{giazotto2006,pothier1997}
\begin{equation}
	T^* = \sqrt{T^2+\frac{V_\mathrm{ctrl}^2}{4\mathcal{L}}}\label{eqn:wfheat}
\end{equation}
in the middle of the control wire, where $\mathcal{L}$ is the Lorenz number. We assume $\mathcal{L}=\mathcal{L}_0=\pi^2k_\mathrm{B}^2/3e^2$ for graphene at low temperatures and sufficiently far from the Dirac point \cite{crossno2016}. $f_d$ and $f_\mathrm{h}$ are plotted for $eV_\mathrm{ctrl}=\Delta$ in Fig.~\ref{fig:Figure5}(c), along with the equilibrium distribution $f_0$ for the base temperature $T=50~\mathrm{mK}$ of the experiment.

Fig.~\ref{fig:Figure5}(d) illustrates the effect of the control voltage on the spectral supercurrent for a phase difference of $\pi/2$ across the junction and $\tau=0.94$. The distribution functions are the same as in Fig.~\ref{fig:Figure5}(c). Compared to the equilibrium case, the double-step distribution selectively removes the low-energy part, while the hot-electron distribution leads to an overall reduction.

\begin{figure*}
\includegraphics[width=\textwidth]{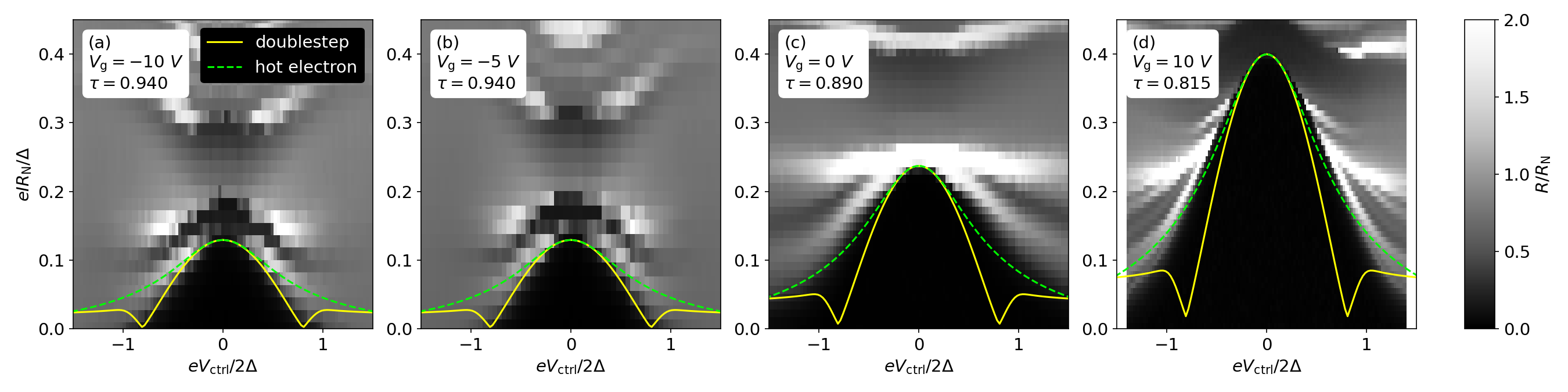}
\caption{Normalized resistance $R/R_\mathrm{N}$ as a function of normalized bias current $eIR_\mathrm{N}/\Delta$ and normalized control voltage $eV_\mathrm{ctrl}/2\Delta$ for different gate voltages $V_\mathrm{g}$. Solid lines are model results for the double-step distribution, dashed lines are for the hot-electron regime (see text for details).}
\label{fig:Figure6}
\end{figure*}

Fig.~\ref{fig:Figure6} shows the comparison of the experimental data to the model for different gate voltages. 
The coupling strength $\tau$ is the only free parameter, which we have adjusted to reproduce the observed critical current at $V_\mathrm{ctrl}=0$. The double-step distribution (solid lines) gives a good description of the data for small control voltage in the p-doped regime. In the n-doped regime, the hot-electron distribution (dashed lines) gives a good description, reflecting the higher power dissipation. Due to the negative spectral supercurrent at high energy, there should be a zero to $\pi$ transition of the Josephson current at $eV_\mathrm{ctrl}/2\Delta\lesssim 1$ for the double-step distribution. This is visible as a node in the model plots, where the critical current vanishes and then reappears. The transition is absent in the hot electron regime. The fact that we observe a monotonic decrease of the critical current without $0-\pi$ transition can be attributed to partial thermalization at higher control voltage even in the p-doped regime. A more detailed model accounting for partial thermalization would require a realistic treatment of Coulomb scattering in the device, which is beyond the scope of our experimental work.

\section{Discussion and perspective}
We have demonstrated that supercurrent can be tuned by a transverse voltage in a ballistic graphene-based Josephson junction. By using this control voltage, we continuously  modify the occupation of the discrete Andreev levels and therefore the magnitude of $I_{\mathrm{c}}$. We highlight two different regimes with our model, i.e., a double-step distribution and a hot electron regime. The first corresponds to the case where the Fermi level, tuned by the gate voltage $V_\mathrm{g}$, sits in the valence band and $I_{\mathrm{c}}$ amplitude is the lowest while the normal state resistance is the highest due to the additional pn-junction formed in the vicinity of the leads by charge transfer (forming an electronic FP cavity and FP interferences are observed via the conductance/resistance oscillations, see Fig.~\ref{fig:Figure2}(a) and the checkerboard pattern when the system is set out-of-equilibrium, see Fig.~\ref{fig:Figure2}(c)). As mentioned before, this double step distribution implies a full heat dissipation by the leads. The second corresponds to the case where the Fermi level sits in the conduction band and $I_{\mathrm{c}}$ amplitude is the highest, while the normal state resistance is the lowest at large charge carrier density, in turn, leading to increased power dissipation and self-heating effect. We note that this is also reflected in the hysteretic behaviour of $I_{\mathrm{c}}$ \cite{courtois2008} (See the asymmetry of $I_{\mathrm{c}}$ the differential resistance map in Fig.~\ref{fig:Figure4}(g) and (h)). While we observe two different regimes, we actually do not observe the sign reversal of the critical current implying the formation of a controllable $\pi$-junction \cite{ryazanov2001,Huang2002} as reported in metallic nanostructures \cite{baselmans1999,baselmans2001} but not in semiconducting systems \cite{schapers2003,schapers2003a}. We have interpreted the absence of $0-\pi$ transition in our samples by the partial charge carrier thermalization at large control voltages. This could be prevented by optimizing the geometry of our system and enlarging the size of the reservoirs. We may suggest that these experiments could be performed by using double gated bilayer graphene for an even more tunable weak link in a Josephson junction \cite{kraft2018}, as the combination of gates does not only allow the control of the charge carrier density and the contact resistance, but also can be used to break the lattice inversion symmetry of the bilayer graphene system, opening a band gap and modifying the band structure itself. This would add another knob to explore many possible configurations of a Josephson junction.

\section*{Acknowledgements}

The authors thank R. M\'elin and C. Winkelmann for fruitful discussions. 
This work was partly supported by Helmholtz
society through program STN and the DFG via the projects
DA 1280/3-1, DA 1280/7-1 and BE 4422/4-1.

\bibliography{Energy_distribution_controlled_JJ} 

\end{document}